\documentclass[twocolumn,showpacs,preprintnumbers,amsmath,amssymb]{revtex4}

\usepackage{graphicx}
\usepackage{dcolumn}
\usepackage{bm}

\newcommand{\be}[1]{\begin{equation}\label{eq:#1}}
\newcommand{\ee}{\end{equation}}
\newcommand{\bea}{\begin{eqnarray}}
\newcommand{\eea}{\end{eqnarray}}

\newcommand{\bt}{\textbf}

\newcommand{\phd}{\phantom{\dag}}
\newcommand{\ph}{\phantom{.}}
\newcommand{\up}{^{\phd}}
\newcommand{\noi}{\noindent}
\newcommand{\no}{\nonumber}

\newcommand{\dvb}[1]{\bm{\nabla}_{\bm{#1}}\up}

\begin{document}
\def\v#1{{\bf #1}}


\title{Magnetic field induced chiral particle-hole condensates}

\author{P. Kotetes}\email{pkotetes@central.ntua.gr}
\author{G. Varelogiannis}
\affiliation{Department of Physics, National Technical University
of Athens, GR-15780 Athens, Greece}

\vskip 1cm
\begin{abstract}
We demonstrate that a chiral particle-hole condensate is always
induced by a number-conserving ground state of non-zero angular
momentum in the presence of a magnetic field. The magnetic
interaction originates from the coupling with the intrinsic
orbital moment of the chiral state when the field is applied
perpendicularly to the plane. According to our numerical results
the induction mechanism is practically temperature independent
providing robustness to these states up to high temperatures. This
opens the door for manipulating the anomalous Hall response
resulting from this intricate class of states for technological
applications while it also suggests that chiral particle-hole
condensates may be hidden in various complex materials.
\end{abstract}

 \maketitle


Among the numerous states that emerge in a strongly correlated
system in the particle-hole channel, the chiral states enjoy
special attention. These states apart from the related symmetry
breaking they effect due to the corresponding pair condensation,
they additionally violate parity and time-reversal because of
their special momentum structure (See e.g. \cite{Volovik}). As a
consequence, the condensate carries an orbital moment that can be
viewed as the effect of a non-zero Berry curvature in
$\bm{k}$-space
\cite{Berry,{LuttingerFerromagnetics},semiclassical}. The arising
orbital moment \cite{OrbitalMagnetization} provides the system
with a magnetic field coupling leading to anomalous Hall transport
and enhanced diamagnetic response. As a matter of fact, such
states have ideal properties for technological applications. In
addition the well established belief that unconventional
particle-hole condensates are hidden in the majority of several
important materials render these states as strong candidates for a
pleiad of unidentified phases. Consequently, it is of high
priority to be in position to generate these states and manipulate
them, at our own disposal.

One of the traditional ways to controllably engineer novel states
of matter is via applying an external field. Typical examples of
field induced states in the area of particle-hole condensates
constitute the field induced charge and spin density waves. There
are several kinds of field induced density waves such as those
occurring in organic quasi-1D conductors \cite{FISDW} or even
confined spin density waves \cite{GV} both generated due to the
orbital coupling with the magnetic field. Nevertheless, one may
also obtain induced density waves driven by the Zeeman coupling
\cite{aperis}.

In this letter we perform a numerical study of the magnetic field induced
planar chiral particle-hole condensates. A similar situation was first 
addressed by R. B. Laughlin \cite{Laughlin} in the context of chiral d-wave superconductors 
to explain the magnetic field induced transition observed in 
Bi$_2$Sr$_2$CaCu$_2$O$_8$, and later extended to the particle-hole channel 
by J.-X. Zhu \textit{et al} \cite{Balatsky}. Since the whole class of
these chiral states are characterized by a universal behaviour we
shall concentrate on a particular state belonging to this class,
the chiral d-density wave state, that recently attracted a lot of attention 
due to its prominent role in explaining the pseudogap regime of the 
cuprates \cite{Tewari,{KVL},{Meissner},{Zhang},{Spin},Partha}. 
By taking into account the effect of an external perpendicular 
magnetic field through its coupling with the intrinsic orbital magnetic
moment of this state, we demonstrate
through a detailed numerical analysis that a chiral state is
necessarily induced or it is strongly enhanced if it already
exists. In addition, we extract the magnetic field dependence of
the two density wave order parameters, which is
different in the above two cases. However, the interplay of the
underlying interaction and the field strength can lead to a
crossover in the magnetic field dependence in the latter case.
Furthermore, we observe that the chiral d-density wave is robust
against the increase of temperature in the presence of the
external magnetic field, a direct consequence of the field driven
enhancement. Our results are striking demonstrating that
\textit{in many materials in which possibly only some components
of a chiral particle-hole condensate develop, there will be
unavoidably an induction of the rest and a concomitant transition
to the complete chiral state in the presence of an external field,
giving rise to the aforementioned intriguing response.}

As we have already mentioned, the representative chiral
particle-hole condensate that we shall consider is the chiral
d-density wave state that constitutes a singlet unconventional
density wave with a planar momentum structure, characterized by
the \textit{commensurate} wave-vector $\bm{Q}=(\pi/a,\pi/a)$. It
is composed by a real $d_{xy}$ charge density wave violating
parity
and an imaginary $d_{x^2-y^2}$ orbital anti-ferromagnetic state
giving rise to local charge currents and zero charge density,
violating time-reversal. This state has been shown to exhibit
unconventional Hall transport and anomalous magnetic response
which is common to any other chiral particle-hole condensate with
zero \cite{SCZ} or finite momentum taking place either in the
singlet or triplet channel of some kind of a spin degree of
freedom, that is characterized by the same Berry curvature.

First of all, the chiral d-density wave is known to give rise to
the Spontaneous Quantum Hall effect \cite{{KVL},{Yakovenko}},
which concerns the generation of a quantized Hall voltage via the
sole application of an electric field. Quite similarly, a
thermoelectric Hall effect can be reproduced by the application of
a finite temperature gradient \cite{Zhang}. Moreover, this state
supports topological spin transport characterized by
dissipationless spin currents in the presence of a Zeeman field
gradient \cite{Spin}. An even more striking behaviour dominates
the magnetic response of this system. The existence of the
intrinsic orbital magnetic moment leads to perfect diamagnetism
and consequently to the Topological Meissner effect demonstrated
in Ref.\cite{Meissner}. In the half-filled case the topological
Meissner effect is identical to the usual superconducting
diamagnetism, and motivated by this feature the chiral d-density
wave state has been proposed to be hidden in the pseudogap phase
of the cuprates. Furthermore, the observed Polar Kerr effect in
YBa$_{2}$Cu$_{3}$O$_{6+x}$ \cite{Xia}, has been considered as a
sharp signature of the chiral d-density wave state in underdoped
cuprates \cite{Tewari}.

To model the chiral d-density wave state we consider the following
Hamiltonian \bea {\cal
H}&=&\sum_{\bm{k}}\left\{\varepsilon(\bm{k})c_{\bm{k}}^{\dag}c_{\bm{k}}\up
+\varepsilon(\bm{k+Q})c_{\bm{k+Q}}^{\dag}c_{\bm{k+Q}}\up\right\}\no\\
&-&\frac{1}{v}\sum_{\bm{k},\bm{k}'}{\cal
V}(\bm{k},\bm{k}')c_{\bm{k}}^{\dag}c_{\bm{k}+\bm{Q}}\up
c_{\bm{k}'+\bm{Q}}^{\dag}c_{\bm{k}'}\up\,,\eea \noi where we have
introduced the single band energy dispersion of the free Bloch
electrons $\varepsilon(\bm{k})=-2t[\cos(k_xa)+\cos( k_ya)]$
arising from the nearest neighbours hopping term and a 4-fermion
interaction driven by an effective separable potential ${\cal
V}(\bm{k},\bm{k}')=V_1f_1(\bm{k})f_1(\bm{k}')+V_2f_2(\bm{k})f_2(\bm{k}')$,
projecting \textit{only} onto the form factors
$f_{1}=\sin(k_xa)\sin(k_ya)$ and $f_2=\cos(k_xa)-\cos(k_ya)$ of
the $d_{xy}$ and $d_{x^2-y^2}$ momentum space orbitals. The
operators $c_{\bm{k}}\up/c_{\bm{k}}^{\dag}$ annihilate/create an
electron of momentum $\bm{k}$ in the reduced Brillouin zone
(R.B.Z.), $v$ is the volume of the system and $a$ the square
lattice constant. For convenience we have omitted the spin
indices, as we are dealing with a spin singlet state. Within a
mean field decoupling we obtain the order parameter of the chiral
d-density wave state
$\Delta(\bm{k})=\Delta_1\sin(k_xa)\sin(k_ya)-i\Delta_2[\cos(k_xa)-\cos(k_ya)]$,
which satisfies the self-consistence equation
\bea\Delta(\bm{k})=-\frac{1}{v}\sum_{\bm{k}'}{\cal
V}(\bm{k},\bm{k}')\left
<c_{\bm{k}'}^{\dag}c_{\bm{k}'+\bm{Q}}\up\right>\,,\eea \noi with
$<>$ denoting thermal and quantum-mechanical average. For
convenience, we adopt a more compact notation by considering the
enlarged spinor $\Psi_{\bm{k}}^{\dag}=(c_{\bm{k}}^{\dag}\phd
c_{\bm{k+Q}}^{\dag})$ and employing the $\bm{\tau}$ Pauli
matrices. The mean field Hamiltonian now becomes ${\cal
H}=\sum_{\bm{k}}\Psi_{\bm{k}}^{\dag}{\cal
H}(\bm{k})\Psi_{\bm{k}}\up=\sum_{\bm{k}}\Psi_{\bm{k}}^{\dag}\ph\bm{g}(\bm{k})
\cdot\bm{\tau}\ph\Psi_{\bm{k}}\up$ where we have introduced the
isovector $\bm{g}(\bm{k})$ defined as $\bm{g}(\bm{k})=\left({\cal
R}e\Delta(\bm{k}),-{\cal
I}m\Delta(\bm{k}),\varepsilon(\bm{k})\right)$. The one-particle
Hamitonian for each $\bm{k}$-mode is a $2\times2$ matrix \bea{\cal
H}(\bm{k})=\left(\begin{array}{cc}
g_3\up(\bm{k})&g_{1}\up(\bm{k})-ig_2\up(\bm{k})\\
g_{1}\up(\bm{k})+ig_2\up(\bm{k})&-g_3\up(\bm{k})\\\end{array}\right)\,.\eea

To incorporate the interaction with the magnetic field we shall
consider only the orbital coupling and neglect the Zeeman term as
it is usually negligible in the case we are considering. To
calculate the orbital moment one has to determine the Berry phase
emerging in this chiral state when $\bm{k}$ changes adiabatically
along a closed loop. For instance, the variation of $\bm{k}$ can
be enforced by the minimal coupling
$\bm{k}\rightarrow\bm{k}-e\bm{{\cal E}}t$ if we apply a constant
electric field $\bm{{\cal E}}$. In this case, the Hamiltonian
becomes parametric ${\cal H}(\bm{k})\rightarrow{\cal
H}(\bm{k},t)$. Within the adiabatic approximation, the emerging
Berry phase can be determined using the instantaneous (snapshot)
eigenstates of the parametric Hamiltonian $|\Phi_{\nu}(\bm{k},t)$,
satisfying ${\cal H}(\bm{k},t)|\Phi_{\nu}(\bm{k},t)\rangle=
E_{\nu}(\bm{k},t)|\Phi_{\nu}(\bm{k},t)\rangle$. In this equation,
$t$, is introduced only as a parameter. This means that these
snapshot eigenstates are not really time-dependent but only
parameter dependent, which in our case coincides with $t$. Our two
band system, is characterized by the snapshot eigenstates
$|\Phi_{\pm}\rangle$ and the corresponding eigenenergies
$E_{\pm}(\bm{k},t)=\pm |g(\bm{k},t)|$. By defining
$g_1\up(\bm{k})=E(\bm{k})\sin\theta(\bm{k})\cos\varphi(\bm{k})$,
$g_2\up(\bm{k})=E(\bm{k})\sin\theta(\bm{k})\sin\varphi(\bm{k})$
and $g_3\up(\bm{k})=E(\bm{k})\cos\theta(\bm{k})$, we obtain a
convenient expression for the snapshot eigenstates of the system
\bea|\Phi_{+}\up(\bm{k},t)\rangle=\left(\begin{array}{l}\cos\left(\frac{\theta(\bm{k},t)}{2}\right),\phantom{-}
\sin\left(\frac{\theta(\bm{k},t)}{2}\right)
e^{i\varphi(\bm{k},t)}\end{array}\right)^T,&\no\\
|\Phi_{-}\up(\bm{k},t)\rangle=\left(\begin{array}{l}\sin\left(\frac{\theta(\bm{k},t)}{2}\right),
-\cos\left(\frac{\theta(\bm{k},t)}{2}\right)
e^{i\varphi(\bm{k},t)}\end{array}\right)^T\,,&\eea

\noi with $E(\bm{k})=|g(\bm{k})|$ and $^T$ denoting matrix
transposition. The orbital magnetic moment
$\bm{m}_{\nu}\up(\bm{k})$ is determined by the relation
$\bm{m}_{\nu}\up(\bm{k})=\frac{e}{2\hbar
i}\langle\dvb{k}\Phi_{\nu}\up(\bm{k})\mid\times\left[{\cal
H}(\bm{k})-E_{\nu}\up(\bm{k})\right]\mid\dvb{k}\Phi_{\nu}\up(\bm{k})\rangle=\frac{e\nu}{\hbar}
E(\bm{k})\bm{\Omega}_{\nu}\up(\bm{k})$ where we have introduced
the Berry curvature
\cite{Spin}\bea\bm{\Omega}_{\nu}\up(\bm{k})=-\frac{\nu
}{2E^3(\bm{k})}\ph\bm{g}(\bm{k})\cdot\left(\frac{\partial\bm{g}(\bm{k})}{\partial
k_x\up}\times\frac{\partial\bm{g}(\bm{k})}{\partial
k_y\up}\right)=\Omega^z_{\nu}\ph\bm{\hat{z}}.\,\eea \noi We notice
the Berry curvature and the orbital moment lie along the $z$-axis,
as a direct consequence of the planar character of our system. The
presence of a perpendicular magnetic field $\bm{{\cal B}}$ enters
the band dispersions of the system in the following way
$E_{\nu}^{\bm{{\cal
B}}}(\bm{k})=E_{\nu}\up(\bm{k})-\bm{m}_{\nu}\up(\bm{k})\cdot\bm{{\cal
B}}$.

Having obtained the energy dispersions of the system in presence
of the magnetic field we may now extract the self-consistence
equations of the $d_{xy}$ and $d_{x^2-y^2}$ order parameters
$\Delta_1,\Delta_2$. The free energy functional is defined as
\cite{OrbitalMagnetization} \bea{\cal
F}=\frac{\Delta^2_{1}}{V_1\up}+\frac{\Delta^{2}_{2}}{V_2\up}
-\frac{1}{v\beta}\sum_{\bm{k},\nu=\pm}Ln\left(1+e^{-\beta
E_{\nu}^{\bm{{\cal B}}}(\bm{k})}\right)\ph\,.\eea \noi with
$\beta=1/k_B\up T$. The first two terms in the free energy
originate from the mean field decoupling and correspond to the
elastic energy waisted for building up the two density wave gaps.
Minimization of this functional with respect to the order
parameter doublet leads to the following system of coupled
self-consistence equations
\begin{widetext} \bea\frac{\partial{\cal
F}}{\partial
\Delta_i\up}=0\Rightarrow\Delta_i\up=\frac{2}{v}\sum_{\bm{k},\nu}\left\{-
V_i\up\Delta_i\up\frac{f_i^2(\bm{k})}{2E(\bm{k})}\nu+\frac{ea^2t}{\hbar}{\cal
B}_z\up V_i\up\widetilde{\Delta}_{i}\up
\frac{s(\bm{k})}{E^2(\bm{k})}\left[1-2\left(\frac{\Delta_i\up(\bm{k})}{E(\bm{k})}\right)^2\right]\right\}n_F\up\left[E_{\nu}^{\bm{{\cal
B}}}(\bm{k})\right],\phd i=1,2\,,\label{eq:sq}\eea
\end{widetext}

\noi that will be used for determining the chiral d-density wave
gaps numerically. In the above, we have introduced
$s(\bm{k})=\sin^2(k_xa)\cos^2(k_ya)+\sin^2(k_ya)$,
$\widetilde{\Delta}_{1,2}=\Delta_{2,1}$, $n_F\up(E)$ the
Fermi-Dirac distribution and multiplied by a factor of two in
order to take into account the electron spin. The first term
corresponds to the equation that we obtain in a zero magnetic
field with the only difference that the energy bands $E_{\nu}\up$
are shifted by the orbital coupling
$-\bm{m}_{\nu}\up\cdot\bm{{\cal B}}$. On the other hand, the
second term is attributed entirely to the magnetic field
interaction with the chiral d-density wave state. As a matter of
fact, it is the essential ingredient for a field induced chiral
d-density wave. To understand how a magnetic field stabilizes this
chiral state we consider the case of $\Delta_1=0$ and
$\Delta_2\neq 0$. In this case the initial state is an orbital
anti-ferromagnet. We may calculate the induced $\Delta_1$
component by setting $\Delta_1=0$ on the right hand side of
Eq.(\ref{eq:sq}). It is straightforward to obtain that
$\Delta_1^{induced}=\frac{ea^2t}{\hbar}{\cal B}_z\up
V_2\up\Delta_2\up{\cal I}({\cal B}_z\up)$ with ${\cal I}({\cal
B}_z\up)$ corresponding to a suitable sum over R.B.Z. of the
remaining terms. Apart from the additional weak dependence on the
field in ${\cal I}$, this formula also agrees with the one derived
in Ref.\cite{Balatsky} using a different method.\textit{ According
to our result a chiral d-density wave state is always generated
even for an arbitrary small magnetic field}.

We now proceed in solving numerically the system of
self-consistence equations. For all our numerical simulations we
set $t=250$meV, $a=5${\AA} and $V_2=150$meV. The last condition
aims to establish a $d_{x^2-y^2}$ density wave of approximately a
$\Delta_2\simeq53$meV gap for all the possible temperatures and
magnetic fields considered here. The latter consideration helps us
to focus solely on the behaviour of the $d_{xy}$ component. For
the calculations we have set up an $128\times128$ grid in the
right upper quadrant of the Brillouin zone. A large number of
points are needed in order to stabilize a value for the $d_{xy}$
order parameter.

\begin{figure}[t]\centering
\includegraphics[width=0.48\textwidth]{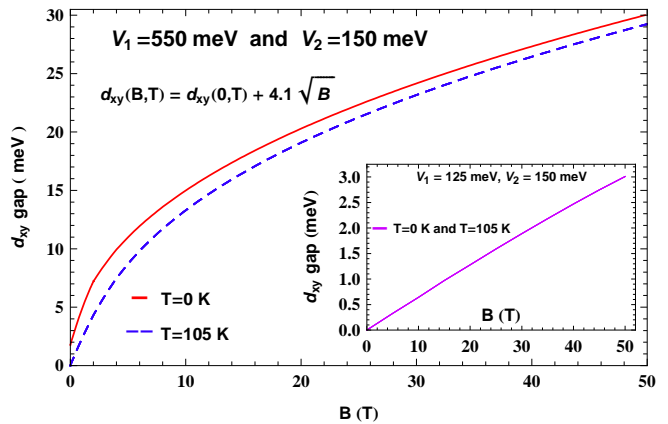}
\caption{(Color online) Magnetic field induced chiral d-density
wave. In the inset, we present the magnetic field dependence of 
the induced $d_{xy}$ component when only a $d_{x^2-y^2}$ already 
exists, for two different temperatures. In the main panel we
demonstrate the strengthening due to the field, of an initially
subdominant $d_{xy}$ component of a preexisting chiral d-density
wave state. The two curves presented for the $d_{xy}$ part differ
only on the zero field gap value arising from the usual BCS
temperature dependence. As we may observe, the induced part of the
chiral state is always temperature insensitive, while the field
dependence changes from linear to square root.}\label{fig:a}
\end{figure}

First of all, we verify numerically the linear dependence on the
magnetic field of the induced $d_{xy}$ order parameter. For the
illustration we consider $V_1=125$meV, a value that in zero
magnetic field would provide a practically zero $d_{xy}$ gap. In
the inset of Fig.\ref{fig:a} we do observe the expected linear
scaling while we also notice the extremely weak temperature
dependence. The latter is attributed to the manner in which the
Fermi-Dirac occupation numbers enter the second part of the
self-consistence equations. Specifically, the occupation numbers
of the two bands add up giving
$n_F\up\left[-E(\bm{k})-m_z\up(\bm{k}){\cal B}_z\up\right]+
n_F\up\left[E(\bm{k})-m_z\up(\bm{k}){\cal B}_z\up\right]\simeq 1$
due to the negligibleness of the magnetic coupling compared to
$E(\bm{k})$.

More intriguing results emerge in the case of an initially
existing chiral d-density wave state. We consider the combination
of potentials $V_1=550$meV and $V_2=150$meV, generating the gaps
$\Delta_1\simeq 1.8$meV and $\Delta_2\simeq 53$meV in the absence
of the field. When the orbital magnetic coupling is switched on,
the $d_{x^2-y^2}$ gap remains mostly unaffected while the $d_{xy}$
component is significantly enhanced. Specifically, the magnitude
of the gap becomes about $15$ times bigger for ${\cal
B}_z\up$=50T. Moreover, we observe that the magnetic field
dependence of the latter gap turns out to be square root contrary
to the linear dependence obtained earlier. Of course, this
difference arises from the first term of the self-consistence
equation which dominates the zero field limit and is fully active
here, compared to the previous situation.

As far as the temperature dependence is concerned, we notice in
the main panel of Fig.\ref{fig:a} that for two totally different
temperature regimes we only have a shift of the two curves, equal
to the difference of the $d_{xy}$ gap value obtained in these
temperatures in the absence of the field. Apparently, the induced
part of the $d_{xy}$ gap is once again temperature independent as
it originates solely from the magnetic field coupling term. We
also obtain the temperature evolution for different magnetic field
values. For example a $\Delta_1=1.8$meV gap initially disappearing
at about $30$K is now robust over a temperature range of more than
$120$K. This is natural if we take into account that the magnetic
field has strengthened the zero temperature gap of the order
parameter that would now collapse to a higher temperature
following the usual BCS behaviour.

\begin{figure}[t]\centering
\includegraphics[width=0.48\textwidth]{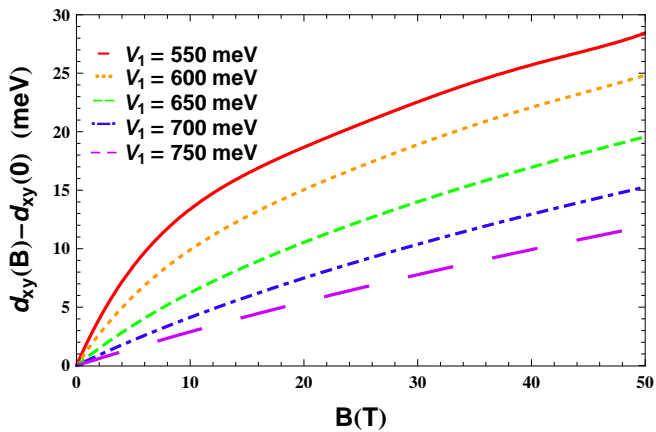}
\caption{(Color online) Magnetic field dependence of the induced
part of the $d_{xy}$ component in a preexisting chiral d-density
wave state for several interaction potentials. For small values of
the potential we obtain the expected square root behaviour, while
after a critical value a crossover to a linear dependence is
triggered. The latter, signals the dominance of the magnetic field
coupling over the zero field 4-fermion $d_{xy}$-channel interaction strength. 
(For the calculations we have used $V_2$=150meV.)}\label{fig:b}
\end{figure}

Finally, we examine the influence of the $d_{xy}$ interaction
potential, on the corresponding gap magnetic field dependence.
Fig.\ref{fig:b} shows the field dependence of the $d_{xy}$ gaps
for several potentials after subtracting the zero field
contribution. Focusing on the induced part of the gap we conclude
that the increase of the interaction strength softens the square
root field dependence turning it into a linear one. This crossover
behaviour is indicative of the enhancement of the magnetic field
coupling that favours a linear trend. As a matter of fact, this
features serve as a potential diagnostic method for the
interaction energy scale.

In conclusion, we have studied numerically the occurrence of
magnetic field induced chiral particle-hole condensates due to the
coupling of their intrinsic orbital moment to a perpendicular
magnetic field. According to our results, if an incomplete chiral
state develops in zero field, such as a $d_{x^2-y^2}$ density
wave, then a $d_{xy}$ is generated spontaneously when the magnetic
interaction is triggered. The magnetic field dependence of the
$d_{xy}$ gap is linear. On the other hand, if a small $d_{xy}$
compared to the $d_{x^2-y^2}$ already exists, then it will have a
square root dependence on the  magnetic field that can become
linear if the interaction potential in the $d_{xy}$ channel
exceeds a critical value. In both cases we obtain a negligible
dependence of the induced order parameters on temperature.
Consequently, chiral particle-hole condensates can survive up to
high temperatures as long as the magnetic coupling persists.

The unavoidable transition to a chiral state has strong impact on
materials that are proposed to host orbital anti-ferromagnetic
states. For instance, the well established d-density wave scenario
\cite{Chakravarty} in the pseudogap regime of the cuprates
provides a unique occasion for the realization of a chiral state,
the chiral d-density wave. In this case, its presence is ensured
when an external perpendicular field is applied but it could also
be present in zero field if the necessary orbital interaction
originates from intrinsic sources such as magnetic impurities.

We are grateful to A. Aperis and S. Kourtis for enlighte\-ning
discussions on numerical methods. Moreover, P.K. is indebted to
Professor C. Panagopoulos and Partha Goswami for valuable comments
and suggestions. One of the authors (P.K.) acknowledges financial
support by the Greek National Technical University Scholarships
Foundation.




\begin{thebibliography}{00}



\bibitem{Volovik} G. E. Volovik, \textit{The Universe in a Helium Droplet}, Oxford Science Publications (2003).

\bibitem{Berry}M.V. Berry, Proc. R. Soc. London A \bt{392}, 45
(1984); R. Resta, Rev. of Mod. Phys. \bt{66}, 899 (1994).

\bibitem{semiclassical}M.-C. Chang and Q. Niu, Phys. Rev. B \bt{53}, 7010
(1996); G. Sundaram and Q. Niu, Phys. Rev. B \bt{59}, 14915
(1999).

\bibitem{LuttingerFerromagnetics}R. Karplus and J.M. Luttinger, Phys. Rev. \bt{95}, 1154
(1954); J.M. Luttinger, Phys. Rev. \bt{112}, 739 (1958); T.
Jungwirth, Q. Niu, and A. H. MacDonald, Phys. Rev. Lett. \bt{88},
207208 (2002); M. Onoda and N. Nagaosa, J. Phys. Soc. Jpn.
\bt{71}, 19 (2002).

\bibitem{OrbitalMagnetization}
J. Shi, G. Vignale, D. Xiao and Q. Niu, Phys. Rev. Lett. \bt{99},
197202 (2007); T. Thonhauser, D. Ceresoli, D. Vanderbilt, and R.
Resta, Phys. Rev. Lett. \bt{95}, 137205 (2005).

\bibitem{FISDW}L. P. Gorkov and A. G. Lebed , J. Phys. Lett. \bt{45}, L433
(1984); M. H\'{e}ritier, G. Montambaux and P. Lederer, J. Phys.
Lett. \bt{45}, L943 (1984); V. M. Yakovenko, Phys. Rev. B.
\bt{43}, 11353 (1991).

\bibitem{GV}G. Varelogiannis and M. H\'{e}ritier, J. Phys.:
Condens. Matter \bt{15}, L673 (2003).

\bibitem{aperis}A. Aperis, M. Georgiou, G. Roumpos, S. Tsonis, G.
Varelogiannis and P. B. Littlewood, Europhys. Lett., \bt{83},
67008 (2008).

\bibitem{Laughlin}R. B. Laughlin, Phys. Rev. Lett. \bt{80}, 5188 (1998) 

\bibitem{Balatsky}Jian-Xin Zhu and A. V. Balatsky, Phys. Rev. B \bt{65}, 132502 (2002).

\bibitem{Tewari} S. Tewari, C. Zhang, V. M. Yakovenko, and S. Das Sarma, Phys. Rev. Lett. \bt{100}, 217004 (2008).
\bibitem{KVL}P. Kotetes and G. Varelogiannis, Europhys. Lett., \bt{84}, 37012 (2008).
\bibitem{Meissner}P. Kotetes and G. Varelogiannis, Phys. Rev. B., \bt{78}, 220509(R) (2008).
\bibitem{Zhang}C. Zhang, S. Tewari, V. M. Yakovenko, and S. Das
Sarma, Phys. Rev. B., \bt{78}, 174508 (2008).
\bibitem{Spin}P. Kotetes and G. Varelogiannis, J. Supercond. Nov. Magn., \bt{22}, 141-145 (2009).
\bibitem{Partha}P. Goswami, arXiv:0905.1533.

\bibitem{SCZ}K. Sun and E. Fradkin, Phys. Rev. B. \bt{78}, 245122
(2008); C. Wu, K. Sun, E. Fradkin, and S.-C. Zhang, Phys. Rev. B.
\bt{75}, 115103 (2007).

\bibitem{Yakovenko}V. M. Yakovenko, Phys. Rev. Lett. \bt{65}, 251 (1990).
\bibitem{Xia} J. Xia, E. Schemm, G. Deutscher, S. A. Kivelson,
D. A. Bonn, W. N. Hardy, R. Liang, W. Siemons, G. Koster, M. M.
Fejer, and A. Kapitulnik, Phys. Rev. Lett. \bt{100}, 127002
(2008).
\bibitem{Chakravarty} S. Chakravarty, R. B. Laughlin, D. K.
Morr, and C. Nayak, Phys. Rev. B \bt{63}, 094503 (2001).










\end{thebibliography}
\end{document}